\documentclass[conference]{IEEEtran}
\IEEEoverridecommandlockouts
\usepackage{cite}
\usepackage{bm}
\usepackage{amsmath,amssymb,amsfonts}
\usepackage{algorithmic}
\usepackage{graphicx}
\usepackage{textcomp}
\usepackage{xcolor}
\usepackage{svg} 
\usepackage{color}
\usepackage{transparent}
\usepackage{graphicx}
\usepackage{import}
\usepackage{booktabs} 
\usepackage{multirow}
\usepackage{siunitx}
\def\BibTeX{{\rm B\kern-.05em{\sc i\kern-.025em b}\kern-.08em
    T\kern-.1667em\lower.7ex\hbox{E}\kern-.125emX}}
\begin{document}

\title{Boosting Physical Layer Black-Box Attacks with Semantic Adversaries in Semantic Communications\\}
	\author{
	\thanks{*Guoshun Nan is the corresponding author}
	\IEEEauthorblockN{
		Zeju Li,
		Xinghan Liu,
		Guoshun Nan*, 
		Jinfei Zhou,
        Xinchen Lyu,
		Qimei Cui,
		Xiaofeng Tao}
	\IEEEauthorblockA{Beijing University of Posts and Telecommunications}
	\IEEEauthorblockA{lizeju@bupt.edu.cn, liuxinghan\_2022@bupt.edu.cn, nanguo2021@bupt.edu.cn, zhouchuluo@bupt.edu.cn, \\lvxinchen@bupt.edu.cn, cuiqimei@bupt.edu.cn, taoxf@bupt.edu.cn}
}
\maketitle

\begin{abstract}
End-to-end semantic communication (ESC) system is able to improve communication efficiency by only transmitting the semantics of the input rather than raw bits. Although promising, ESC has also been shown susceptible to the crafted physical layer adversarial perturbations due to the openness of wireless channels and the sensitivity of neural models. Previous works focus more on the physical layer white-box attacks, while the challenging black-box ones, as more practical adversaries in real-world cases, are still largely under-explored. To this end, we present SemBLK, a novel method that can learn to generate destructive physical layer semantic attacks for an ESC system under the black-box setting, where the adversaries are imperceptible to humans. Specifically, 1) we first introduce a surrogate semantic encoder and train its parameters by exploring a limited number of queries to an existing ESC system. 2) Equipped with such a surrogate encoder, we then propose a novel semantic perturbation generation method to learn to boost the physical layer attacks with semantic adversaries. Experiments on two public datasets show the effectiveness of our proposed SemBLK in attacking the ESC system under the black-box setting. Finally, we provide case studies to visually justify the superiority of our physical layer semantic perturbations. 

\end{abstract}

\begin{IEEEkeywords}
Semantic communications, Physical Layer Attacks, Black-box Attacks, Generative adversarial networks
\end{IEEEkeywords}

\section{Introduction}
The rapid development of mobile applications puts unprecedented pressure on wireless networks, leading to severe traffic congestion and intolerable delays \cite{ITU}. Recently proliferated AI-enabled services, such as IoT (Internet of Things), self-driving and VR/AR, will further exacerbate the above traffic burden in the future due to the predicted data explosion. Meanwhile, existing wireless communications systems, which are built on the Shannon information theory,
primarily focused on how to accurately and effectively transmit raw symbols from the transmitter to the receiver. Along this direction, quantifying the maximum transmission data rate that can be supported by a communication channel gradually pushes the system capacity to the Shannon limit. Such a dilemma motivates us to rethink the design of the wireless communication paradigm beyond the Shannon approach. 


Recent advances in powerful deep learning technologies open up opportunities for compressing the input with neural models and then transmitting the compressed representations of data over wireless channels. Such a way of paradigm is known as end-to-end semantic communication (ESC)\cite{xie2020deep}, which can be considered as the breakthrough beyond the Shannon approaches. Specifically, semantic communications aim at the successful transmission of semantic information conveyed by the input rather than the accurate reception of every single symbol or bit. 
Recently works have discussed the promise of ESC systems for the transmission of text \cite{xie2020deep}, speech \cite{weng2021semantic}, images \cite{yang2022ofdm} and videos \cite{jiang2022wireless}, showing great potentials for the high demand applications in future 6G networks. 

\begin{figure}[t]
	\centering
	\includegraphics[width=90mm]{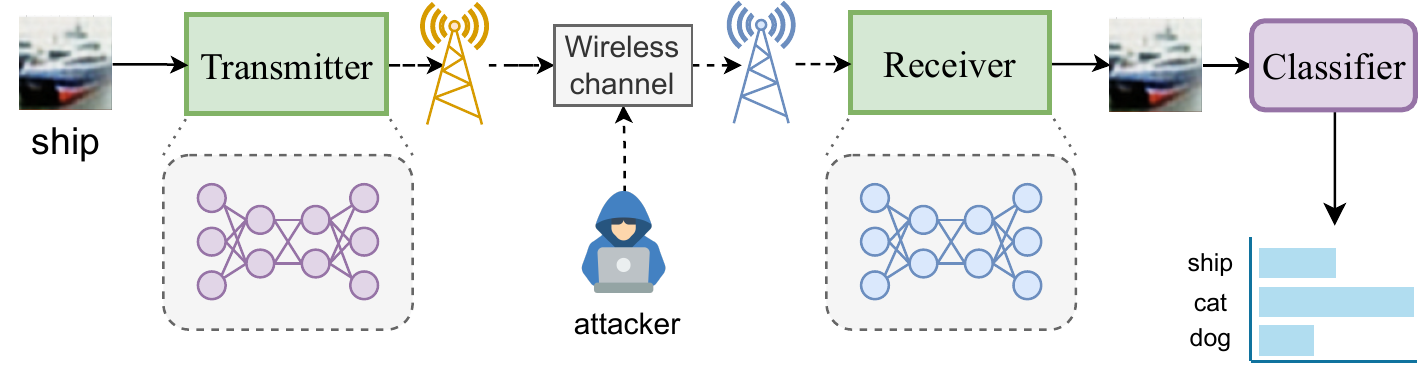}
	\caption{An end-to-end semantic communication system. An attacker generates the physical layer adversarial perturbations and then fools the classifier at the receiver side to make an incorrect semantic decision, i.e., interpreting the ``ship'' type as the ``cat'' type.}
	\label{fig1:fig1}
 \vspace{-4mm}
\end{figure}

Although promising\cite{shi2021semantic,7536171}, ESC has also been shown susceptible to the crafted physical layer adversarial perturbations due to the openness of wireless channels \cite{sadeghi2019physical} and the sensitivity of neural models \cite{baluja2017adversarial}. We give an example to demonstrate the vulnerability of an ESC system as follows. Fig. \ref{fig1:fig1} illustrates an existing ESC system, which mainly consists of three deep learning-based modules, including a transmitter, a receiver and a classifier. The transmitter first extracts the semantics of the input image ``ship'', and then sends the modulated symbols to the wireless channels. The receiver reconstructs the image based on the semantics and then the classifier interprets the semantics as the ``ship'' category. An attacker can send adversarial perturbations to the wireless channel and then may mislead the classifier to make an incorrect decision. As shown in Fig. \ref{fig1:fig1}, the classifier incorrectly interprets the ``ship'' as ``cat'', as the confidence score of the ``cat'' category is the largest among the three candidates.     

\begin{figure*}
	\centering
	\includegraphics[scale=0.6]{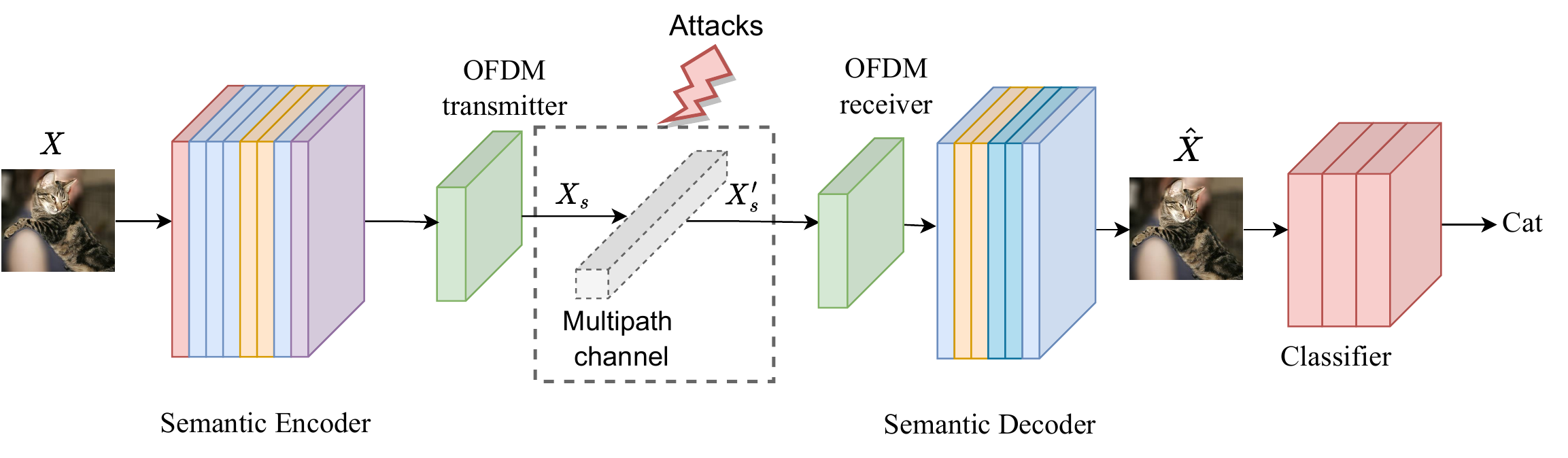}
	\caption{Overview of the end-to-end semantic communication system used in our paper. The system consists of multiple modules including a semantic encoder, an OFDM transmitter, an OFDM receiver, a semantic decoder, and a classifier. An attacker is able to generate malicious perturbation from the wireless channel.}
	\label{fig2:fig2}
 \vspace{-5mm}
\end{figure*}

Previous adversarial attacks in the field of machine learning can be mainly divided into white-box attacks and black-box ones \cite{Explaining_and_harnessing_adversarial_examples,papernot2017practical}. The former attacks have access to the model's parameters, such as FGSM\cite{Explaining_and_harnessing_adversarial_examples} and 
PGD\cite{Towards_deep_learning_models_resistant_to_adversarial_attacks}. 
Conversely, for the latter, we only know the model inputs and we can query to obtain output labels or confidence scores. 

Existing efforts for semantic communications focus more on the white-box attacks
, while the challenging physical layer black-box ones\cite{li2022sembat}, as more practical adversaries in real-world cases, are still largely under-explored. Existing black-box methods in the field of machine learning may not be directly applied to an ESC system due to two reasons: 1) These methods mainly focus on generating black-box adversarial perturbations based on a raw input of a neural model. While in practice, an attacker is unable to access the input data in a communication system and can receive the wireless signal and potentially decode the transmissions. 2) Existing attacks, such as FGSM and PGD, are mainly indiscriminate to all information as physical layer attacks. While the semantics conveyed underlying the data, the central of semantic communications, are largely ignored during the adversarial learning procedure. 

To fill the above gap, this paper proposes SemBLK, a novel method that can learn to generate destructive physical layer semantic attacks for an ESC system under the black-box setting. The two key ingredients of our SemBLK are a surrogate semantic encoder and a novel semantic perturbation generator. Our surrogate encoder learns to output the semantic representations via a limited number of queries to an oracle, and then the perturbation generator can boost the physical layer attacks by learning to craft the semantic adversaries that aim to fool the classifier to make an incorrect interpretation. We conduct experiments to verify the effectiveness of our method. We summarize our contributions as follows:
\begin{itemize}
\item We present SemBLK, a practical physical layer black-box attack method for deep learning-based semantic communication systems, where the adversaries are imperceptible to humans.
\item
We introduce a surrogate semantic encoder to mimic the semantic encoder of the existing ESC system and train its parameters by exploring a limited number of queries to the system, augmenting the training instances with GAN methods. 
\item 
Equipped with our surrogate encoder, we then propose a novel semantic perturbation generation method to learn to boost the physical layer attacks with semantic adversaries.
\item
We conduct extensive experiments on two public benchmarks to show the superiority of our black-box attacks.
\end{itemize}

\begin{figure}
	\centering
	\includegraphics[width=83mm]{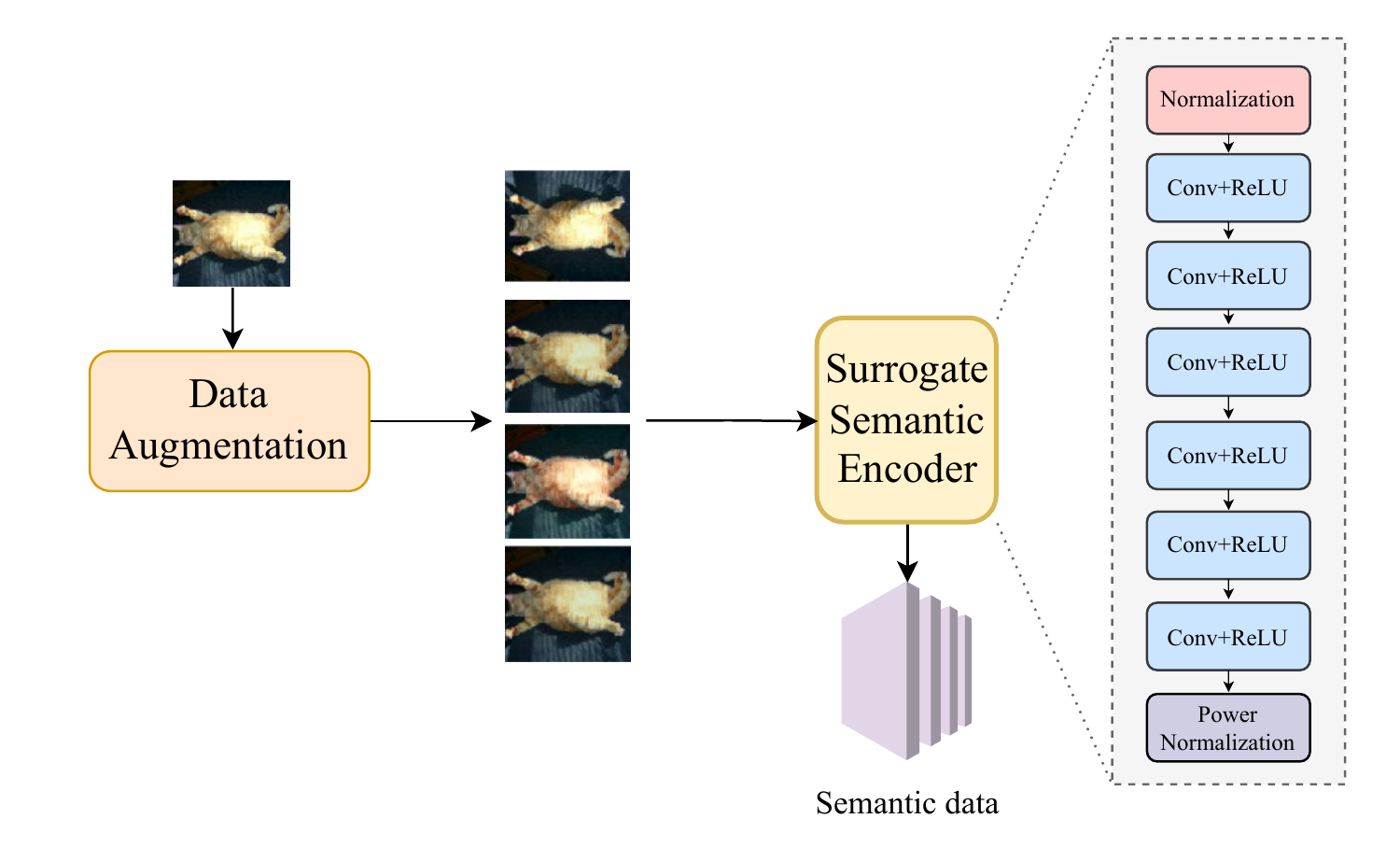}
	\caption{The architecture of our surrogate encoder and the data augmentation procedure to train the encoder. }
	\label{fig4:fig4}
\end{figure}
\begin{figure*}[h]
	\centering
	\includegraphics[scale=0.72]{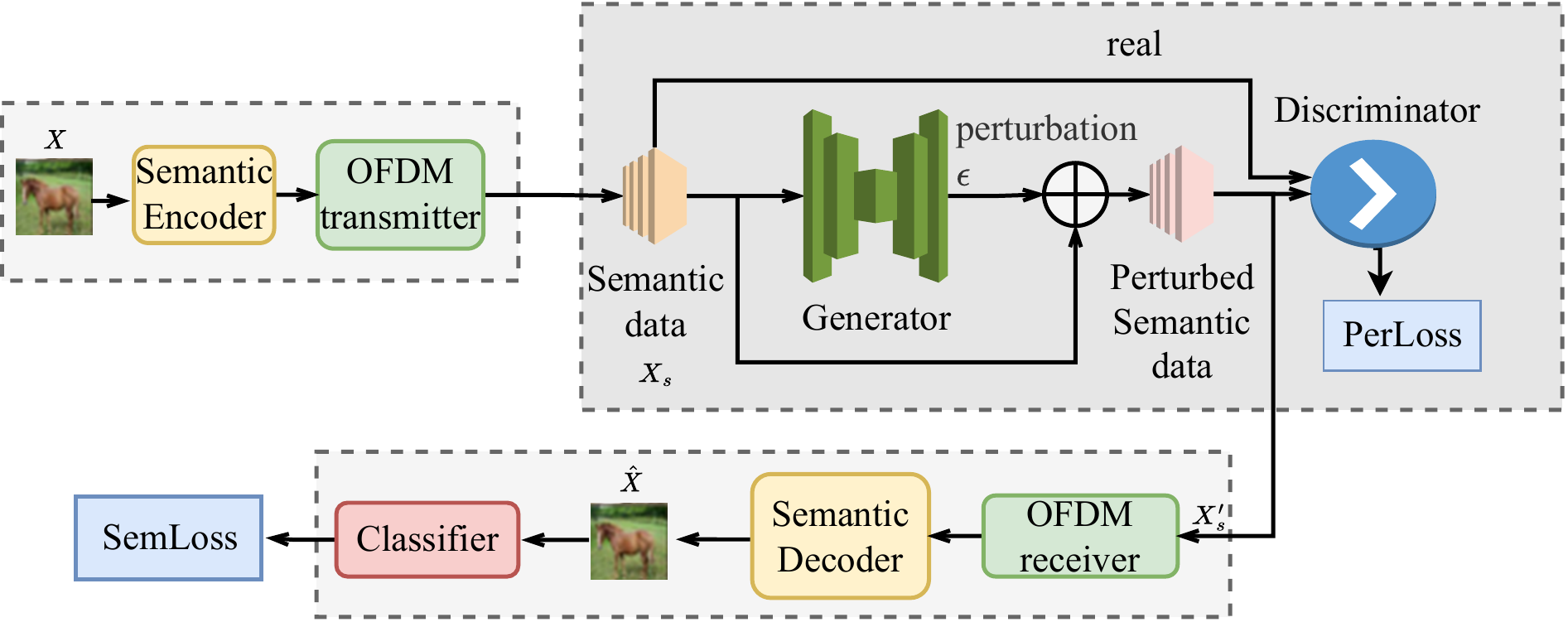}
	\caption{Overview of our perturbation generator. It consists of a generator $\mathcal{G}$ and a discriminator $\mathcal{D}$. The generator $\mathcal{G}$ crafts the adversarial perturbation $\epsilon$. The goal of $\mathcal{D}$ is to distinguish between real semantic symbols $X_s$ and perturbed data $X_s+\epsilon$.}
	\label{fig5:fig5}
\end{figure*}

\section{End-to-end Semantic Communication System}
\label{section:Semantic Communication System}
{
Fig.\ref{fig2:fig2} shows the architecture of an end-to-end semantic communication (ESC) system. We refer to previous JSCC-OFDM \cite{yang2022ofdm} as the backbone and additionally 
introduce a classifier to interpret semantics at the receiver side. We outline each component of the ESC system as follows.

\subsection{Semantic Transmitter}

\noindent
\textbf{Semantic Encoder:}
The semantic encoder uses convolutional and 
residual layers to extract image features.
The convolutional layers extract the input features and 
obtain more detailed semantic feature information.
The residual layers prevent 
performance degradation when the network 
depth is deepening 
and reduce the amount of computation in subsequent layers.\par

\noindent
\textbf{OFDM Transmitter:} 
The OFDM transmitter performs an inverse fast Fourier transform (IFFT) on the data, inserts a cyclic prefix (CP), which protects the signal from interference, and then shears the signal to reduce the peak-to-average power ratio (PAPR).

\subsection{Wireless Channel}
Without loss of generality, we use Rayleigh fading channel in our paper. We parameterize the channel using a discrete channel transfer function:
\begin{equation}
    X_{s} = h(y;\sigma_{0}^{2},...,\sigma_{L-1}^{2},\sigma^{2})= h * X_{s} + w
\end{equation}
where $*$ denotes the convolution operation, $h$ denotes the sample sample space channel impulse response, and $L$ is the number of multipath. $w$ represents the additive Gaussian noise. Each path experiences independent Rayleigh fading satisfying $h_{l} \sim \mathcal{CN}(0,\sigma_{l}^2)$ for $l=0,1,...,L-1$. The power of each path follows $\sigma_{l}^{2}=\alpha_{l}e^{-\frac{1}{\gamma}}$, where $\alpha_{l}$ is a normalization coefficient to satisfy $\sum_{l=0}^{L-1}\sigma_{l}^{2}=1$. $\gamma$ is the time decay constant.

\subsection{Semantic Receiver}

\noindent
\textbf{OFDM Receiver:}
The OFDM receiver removes the CP from the received data, 
performs FFT, channel estimation and channel equalization to obtain the semantic information. \par 

\noindent
\textbf{Semantic Decoder:}
The semantic decoder aims to reconstruct the input image from the received semantic data, and it shares the same architecture as the semantic encoder. 

\noindent
\textbf{Classifier:}
We feed the reconstructed image to a classifier to interpret semantics. Here we use Mobilenetv2\cite{sandler2018mobilenetv2} as the classifier.
}


\section{Our method}
{
We train the ESC system and then treat it as an oracle for the physical layer black-box attacks. Existing black-box attacks require a large number of queries to obtain the confidence scores or labels, while these attacks can be easily detected. Therefore, we introduce a surrogate encoder \cite{papernot2017practical} to mimic the semantic encoder and then train the surrogate with a limited number of queries. Equipped with the surrogate, we also present a perturbation generator that is able to learn to craft the physical layer adversaries for the ESC system. We first delve into the surrogate encoder. 
}

\subsection{Surrogate Semantic Encoder}
We simply use the fully-connected convolutional neural networks (CNN) as our surrogate encoder, since CNN has been proved effectively for image feature extraction. For the existing ESC system. We rely on the surrogate encoder to generate a limited number of queries to obtain the output labels. To train our surrogate encoder with and reduce the number of queries, we augment the input data with a simple yet effective GANs\cite{dos2019data} method. Such a method can not only enrich the training instance but also align the semantic to the original data. Fig.\ref{fig4:fig4} demonstrates the architecture of our surrogate encoder and the augmentation method. Next, we show how we learn to craft the semantic perturbations equipped with such a surrogate encoder. 

\subsection{Adversarial Perturbation Generator}
	\par{
The semantic representations generated by our surrogate encoder will be modulated by the OFDM transmitter and then sent to wireless channels. We denote the OFDM symbols in the channel as $X_s$. Considering that the existing adversarial attack methods cannot be directly applied to our semantic communication system, hence we consider semantic adversaries from the open wireless channel under the black-box setting. 

Fig.\ref{fig5:fig5} shows the architecture of our proposed SemBLK. Our SemBLK consists of a generator $\mathcal{G}$ and a discriminator $\mathcal{D}$. The generator $\mathcal{G}$ crafts the adversarial perturbation $\epsilon$, which will be added to the semantic symbols $X_s$. The goal of $\mathcal{D}$ is to distinguish between real semantic symbols $X_s$ and perturbed data $X_s+\epsilon$. While the generator $\mathcal{G}$ is able to produce adversarial perturbations to distort the semantic information. Next, we show how the perturbation generator learns to craft the adversaries.   
	}
\subsection{Loss Function}
 Equipped with the generator $\mathcal{G}$ and the discriminator $\mathcal{D}$, we can give the perturbation loss as follows.
	\begin{equation}
		L_{Per} = \mathbb{E}_{X_s} \log{\mathcal{D}(X_s)} + \mathbb{E}_{X_s} \log{(1 - \mathcal{D}(\epsilon+\mathcal{G}(X_s))}
	\end{equation}
To guide the training of the generator with the feedback of incorrect semantic interpretations, we also introduce a semantic loss $L_{Sem}$, which can be expressed as follows. 
	\begin{equation}
		L_{Sem}= \mathbb{E}_{X_s} \mathcal{L}(X_s)
	\end{equation}
where $\mathcal{L}$ is the cross-entropy loss for classifications. Then we are able to train the perturbation generator by predicting the type of perturbed image 
	as an incorrect one, which can be considered a multi-task learning procedure. The final loss $L$ can be formulated as follows.
	\begin{equation}
		L=L_{Sem}+wL_{Per}
	\end{equation} 	
where $w$ is a hyper-parameter to indicate the importance of the two losses. The parameters can be optimized by solving the minimax-game as follows. 
	\begin{equation}
		arg\mathop{min}\limits_{\mathcal{G}}\mathop{max}\limits_{\mathcal{D}}{L}
	\end{equation}
 
Considering the constraints on the number of queries for the training, we also use the data generated by our augmentation method to train the surrogate encoder, as shown in Fig.\ref{fig4:fig4}. Finally, we can directly apply the generator to the existing semantic communication system to craft black-box adversarial attacks.

\subsection{Imperceptibility of Our Black-box Attacks}
To be practical in real-world cases, we also consider the imperceptibility of the black-box attacks at the receiver side and meanwhile introduce a threshold to secure the constructed image quality measured by structural similarity
index measure (SSIM)\cite{wang2004image}. Let $X$ and $X'$ denote an input image and the reconstructed one, the above threshold $SSIM(X,\hat{X})$ can be formulated as follows.    
	\begin{equation}
		SSIM(X,\hat{X})=[l(X,\hat{X})]^{\alpha} + [c(X,\hat{X})]^{\beta} + [s(X,\hat{X})]^{\gamma} 
	\end{equation}
	\begin{equation}
		l(X, \hat{X})=\frac{2 \mu_X \mu_{\hat{X}}+c_1}{\mu_X^2+\mu_{\hat{X}}^2+c_1}
	\end{equation}
	\begin{equation}
		c(X, \hat{X})=\frac{2 \sigma_{X\hat{X}}+c_2}{\sigma_X^2+\sigma_{\hat{X}}^2+c_2}
	\end{equation}
	\begin{equation}
		s(X, \hat{X})=\frac{\sigma_{X \hat{X}}+c_3}{\sigma_X \sigma_{\hat{X}}+c_3}
	\end{equation}
where $l(X, \hat{X})$, $c(X, \hat{X})$ and $s(X, \hat{X})$ are the brightness comparison, contrast comparison, and structure comparison, respectively. $\mu_X$ and $\mu_{\hat{X}}$ represent the mean value of $X$ and $\hat{X}$ respectively, $\sigma_X$ and $\sigma_{\hat{X}}$ represent the standard deviation of $X$ and $\hat{X}$ respectively. $\sigma_{X\hat{X}}$ represents the covariance of $X$ and $\hat{X}$. 
And $c_1,c_2,c_3$ are constants to avoid errors caused by denominator 0. To be more practical for the semantic communication system, we can simplify the threshold $SSIM(X,\hat{X})$ as follows:
	\begin{equation}
		SSIM(X,\hat{X})=\frac{\left(2 \mu_X \mu_{\hat{X}}+c_1\right)\left(\sigma_{X \hat{X}}+c_2\right)}{\left(\mu_X^2+\mu_{\hat{X}}^2+c_1\right)\left(\sigma_X^2+\sigma_{\hat{X}}^2+c_2\right)}
	\end{equation}

\section{Experiments}
We conduct experiments on CIFAR10, a popular dataset that consists of 
60,000 32×32-pixel images in 10 classes, with 6,000 images per class. We follow the previous work\cite{yang2022ofdm} to use 50,000 images as training instances and the rest 10,000 ones as test examples. Next, we detail evaluation metrics, baselines, experimental settings, the effect of surrogate model, and give some insightful conclusions based on our observations. 

\subsection{Evaluation Metrics}
We employ multiple metrics to measure the performance of our method, 
including Peak Signal to Noise Ratio (PSNR), 
Structural Similarity
Index Measure (SSIM), Peak to Average Power Ratio (PAPR) and accuracy of classification (ACC). We outline these metrics as follows.
\begin{itemize}
	\item $\bold{PAPR}$ is to measure the transmitting 
performance of OFDM system, if there is a large 
PAPR, which requires a high linear dynamic range
 of the transmitting power amplifier of OFDM system. 
 If the dynamic range of the amplifier cannot meet the requirements, 
 it will cause nonlinear distortion of the output signal, 
 and such a communication system will have no practical significance.
 \item 
$\bold{PSNR}$ is an objective measure of picture quality, 
representing the effect of background noise on image quality, 
the higher the PSNR, the less the picture is affected by noise.
\item 
$\bold{SSIM}$ is used to represent the similarity of 
the original image and the received image, compared to PSNR is more in line with the intuitive feeling of the human eye on the image.
\item
$\bold{ACC}$ is the accuracy of the classification 
result of the classifier after semantic decoder,
 which indicates the effectiveness of the guaranteed 
 upper layer service.

\end{itemize}

\subsection{Baselines}
We introduce three existing attacks, including FGSM\cite{Explaining_and_harnessing_adversarial_examples}, PGD\cite{Towards_deep_learning_models_resistant_to_adversarial_attacks} and ATN \cite{baluja2017adversarial}, as baselines to compare with our approach. The descriptions for these attacks are given
as follows.
\begin{itemize}
	\item \textbf{FGSM} adds perturbation to the image along the gradient direction, 
	which increases the loss function and leads the model to get wrong classification results.
	\item \textbf{PGD} takes FGSM a step further by 
	using multiple iterations to add perturbations along the gradient direction on the image.
	\item \textbf{ATN} trains a deep neural network that converts the original image into an adversarial sample.
 \end{itemize}

\subsection{Experimental Settings}
We use Pytorch to implement the neural 
network blocks and the OFDM communication model. 
We utilize Adam \cite{kingma2014adam} to train our surrogate encoder, generator $\mathcal{G}$ and discriminator $\mathcal{D}$ in SemBLK. Moreover, we set hyper-parameter $w$ as 0.1. Attackers don't have knowledge of the semantic communication systems under the black-box setting. We split the training set into two parts. 40,000 images are used to train the targeting ESC system and the rest of the 10,000 ones are used to train the surrogate encoder. We also use 1,000 images in the overall test set of 10,000 images for the evaluation of our surrogate encoder. We use Rayleigh fading channel and set SNR as 10 in the simulation. 
 The initial learning rates 
 of all networks are set as 0.0005, and these rates gradually
  decreased to zero as the number of iterations increases.
   Our model is trained by Cuda of NVIDIA GeForce RTX3090 GPU.
\subsection{Effectiveness of Surrogate Model}\label{AA}

\begin{table}[htbp] 
	\begin{center}   
		\caption{Effectiveness of our surrogate model}  
		\label{table:surrogate} 
		\begin{tabular}{ccccc}
			
			\toprule 
			&  PSNR  & SSIM & PAPR & ACC \\ 
			\midrule 
			Original ESC&	 25.23 &  0.85 & 11.58 & 0.82 \\ 
			\textbf{ESC with surrogate}&	 24.27 &  0.83 & 11.91 & 0.79 \\  
   \midrule
			ESC with surrogate+RandAugment&	 22.61 &  0.82 & 11.39 & 0.78 \\
			\textbf{ESC with surrogate+GANs}&	 \textbf{23.47} &  \textbf{0.81} & \textbf{12.32} & \textbf{0.80} \\ 
			\bottomrule 
		\end{tabular}
	\end{center}
\end{table}
The first two rows of Table \ref{table:surrogate} show that the ESC system equipped with our surrogate model is able to achieve comparable results to the oracle under the same test set in terms of PSNR, SSIM, PAPR and ACC. These results confirm the effectiveness of our substitute model. As there will be very few instances to train the surrogate encoder in a real-world scenario, we build a small sub-dataset from 10,000 training instances of the surrogate encoder and augment these instances with existing approaches \cite{cubuk2020randaugment} and \cite{dos2019data}. The sub-dataset includes 1,000 images, with 100 images per category. We finally generated 9,000 more images, with 1,000 images in each category after data augmentation.

			
			  

The last two rows of Table \ref{table:surrogate} report the comparisons of two augmentation methods on the sub-dataset, i.e, the GAN approach \cite{dos2019data} and RandomAugment approach \cite{cubuk2020randaugment}. We observe that the performance of our surrogate encoder trained by the GAN approach outperforms the ones by the RandomAugment approach. We also find that our surrogate encoder is able to achieve comparable results by exploring only 1,000 original training instances. Hence, equipped with such an augmentation method, we can significantly reduce the number of queries to an existing ESC system, which is practical in real-world cases.

\subsection{The results of attacks}

As shown in Table \ref{table:duikangyangben}, we use these methods to 
make adversarial samples to attack the sender's source input data and compare the 
experimental results. We can find that PGD 
has the best attack effect, but the 
attack effect of our proposed SemBLK is close to 
the effect of FGSM, which shows that SemBLK can also 
degrade the performance of our semantic communication system.
Adversarial attacks on the source input images verify the effectiveness and feasibility of our method. However, adversarial attacks on the input images are an ideal situation. While in practice, it is difficult for an attacker to obtain and control the sender's source input data.
Hence, we consider adding perturbations 
 to semantic information at the physical channel layer, so as to 
 conduct efficient and imperceptible black-box attacks on semantic communication
  systems.
  \begin{table}[h] 
  	\begin{center}   
  		\caption{The attack effect of generating adversarial 
  			examples by FGSM, PGD, ATN, SemBLK}  
  		\label{table:duikangyangben} 
  		\resizebox{80mm}{16mm}{
  		\begin{tabular}{ccccc}
  			
  			\toprule 
			&PSNR & SSIM  &  PAPR & ACC \\ 
			\midrule 
                 No Attack   &25.23 &  0.85 & 11.58 & 0.82 \\
			FGSM & 23.28 & 0.78 & 11.28  &0.14 \\ 
			PGD & 24.10 & 0.82 & 11.57 & 0.06 \\ 
			ATN & 23.45& 0.78 &11.69 &   0.28 \\
			SemBLK & 23.22 & 0.79 &11.54 & 0.09 \\
  			\bottomrule 
  		\end{tabular}
  	}
  	\end{center}
  \end{table}


\begin{figure}
	\centering
	\includegraphics[width=78mm]{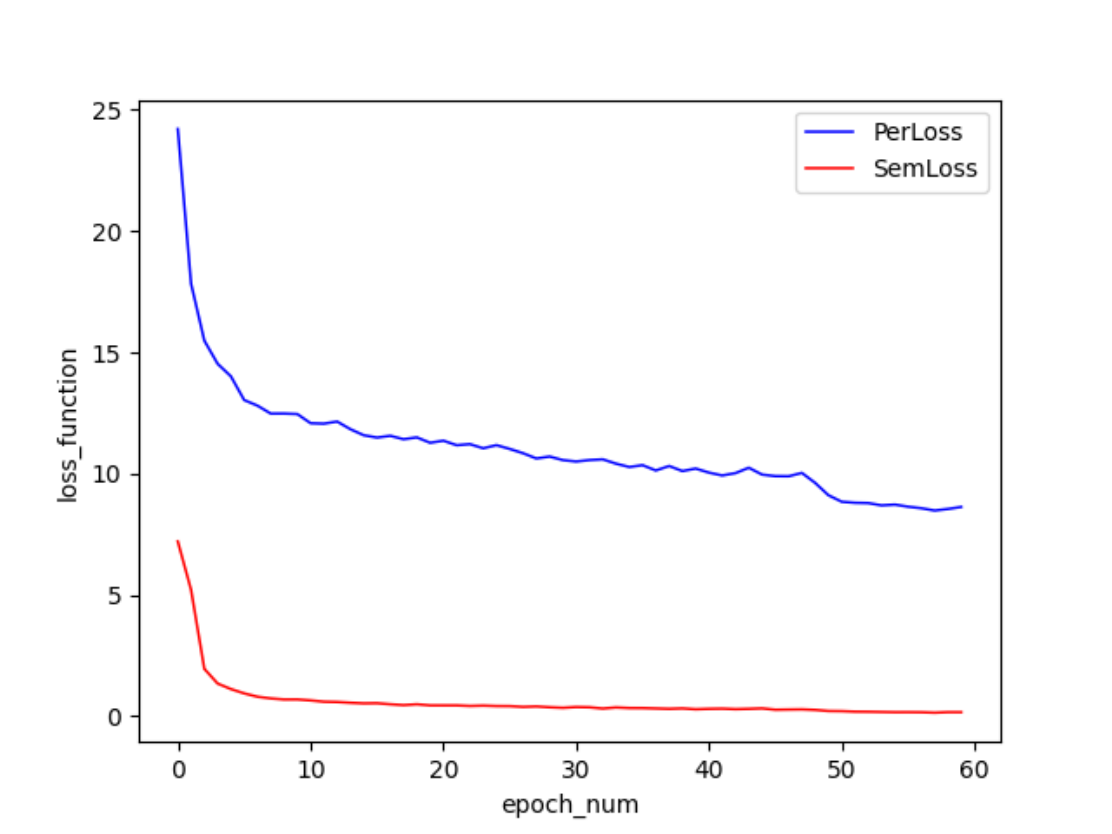}
	\caption{The training curves for PerLoss and SemLoss. We can observe that as the number of iterations increases, SemLoss and PerLoss can gradually decrease and the hence the generator can be optimized step by step. }
	\label{fig6:fig6}
\end{figure}

We use FGSM, PGD and ATN to attack the physical channel 
layer and record the results. 
The four methods add perturbations 
to the semantic information and 
feed the perturbed semantic data as input to the OFDM receiver and the semantic decoder.
In table \ref{table:channel}, we can observe that our proposed SemBLK 
attacks significantly outperform the previous methods FGSM and PGD. For example, the semantic interpretation accuracy of the ESC system under four attacks FGSM, PGD, ATN and SemBLK are $0.66$, $0.75$,   $0.72$ and $0.49$, respectively. However, PSNR, PAPR and SSIM under our SemBLK attacks slightly dropped. These results indicate that our black-box attacks are destructive yet imperceptive.

 \begin{table}[h] 
	\begin{center}   
		\caption{The attack effect of
			 generating perturbations 
			  and overlaying on the semantic data 
			  in physical channels by FGSM, PGD, ATN, SemBLK}  
		\label{table:channel} 
		
		\resizebox{85mm}{16mm}{
		
		\begin{tabular}{ccccc}
			
			\toprule 
			&PSNR & SSIM  &  PAPR & ACC \\ 
			\midrule 
                  No Attack   &25.23 &  0.85 & 11.58 & 0.82 \\
			FGSM & 21.06 & 0.70 & 11.69  &0.66 \\ 
			PGD & 22.08	 & 0.77 & 11.31 & 0.75 \\ 
			ATN & 23.14& 0.78 &11.37 &   0.72 \\
			\textbf{SemBLK} & \textbf{22.91} &  \textbf{0.73} &\textbf{12.11} & \textbf{0.49} \\
			\bottomrule 
		\end{tabular}
	}

	\end{center}
\end{table}

\par{The experimental results show that the effect of using a surrogate encoder
 demonstrates a high PAPR,
 which may cause the transmitter to transmit less effectively.
This is a drawback associated with the attack through the surrogate model, 
but it can substantially reduce the correct classification rate of the upper layer services 
with less impact on the images.
Although our attack causes a decrease in PSNR, 
SemBLK has no major difference in this evaluation metric compared to several other attacks; 
the SSIM does not change much compared to the preattack, 
which is more indicative of the stealthiness of the attack.}
\begin{figure}
	\centering
	\includegraphics[width=78mm]{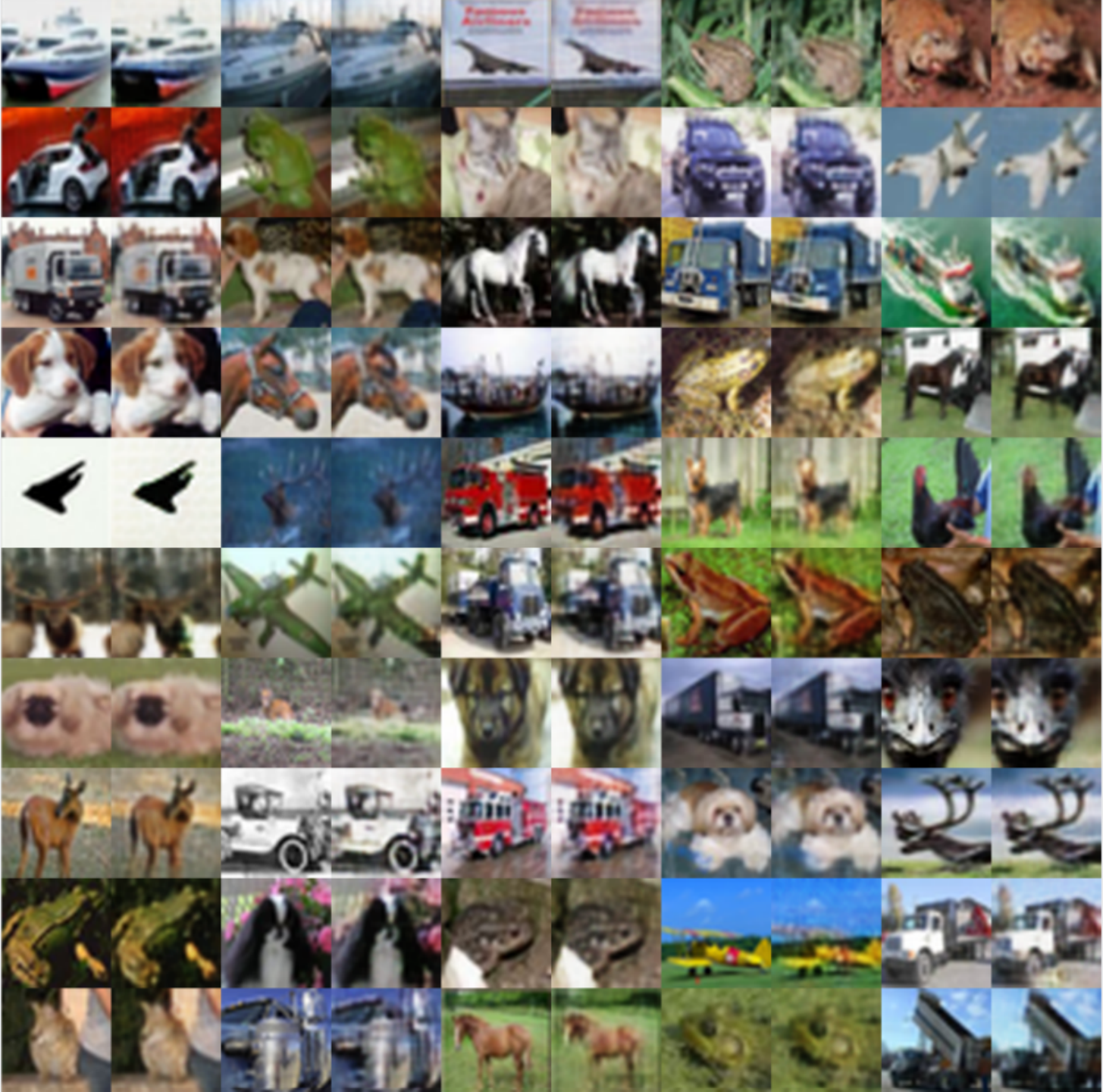}
	\caption{Comparisons of the input images and reconstructed images of 
in semantic communication systems under our SemBLK attacks. The comparisons show that our black-box attacks are imperceptive to  humans.}
	\label{fig7:fig7}
\end{figure}
\subsection{Case study}
Fig.\ref{fig7:fig7} shows 50 cases selected from CIFAR10 to visually demonstrate that the adversaries generated by SemBLK are imperceptive to our humans, while these black-box attacks are able to significantly degrade the semantic interpretations at the receiver side. For each pair of images in Fig. \ref{fig7:fig7}, the left one is an image constructed by the receiver of the ESC system without our attack, and the right one refers to the image transmitted under the physical layer semantic perturbations. We observe that the two images visually appear quite similar to each other with our human eyes, while the accuracy of semantic interpretation of right images, central of the semantic communications, is less than 31 percent on average, compared with the accuracy of the left images without attack. These cases further justify the superiority of our black-box semantic adversaries at the beginning. 

\section{Conclusion}
This paper presents SemBLK, a novel method that aims 
to generate physical layer black-box adversarial attacks 
for end-to-end semantic communication systems. We first use a limited number 
of queries, as well as data augmentation methods
 to train a substitute semantic encoder, and then generate adversarial perturbations that are able to mislead
 semantic interpretation at the receiver. In experiments, we observe that the black-box attacks generated by our SemBLK can significantly degrade the semantic communication system, while these adversaries generated are imperceptive to our humans. We believe our work provides some useful insights into the physical-layer robustness of semantic communications. In the future, we will deploy our proposed method to real scenarios and more semantic communication systems.
\section*{Acknowledgment}
This work was supported in part by National Natural Science Foundation of China (61971066), and in part by the research foundation of Ministry of Education and China Mobile under Grant MCM20180101.
\bibliographystyle{IEEEtran}
\bibliography{conference_101719}
\end{document}